\documentclass[12pt]{article}
\pdfoutput=1
\usepackage{amsmath,amssymb}
\usepackage{graphicx}
\usepackage{subfigure}
\usepackage[pdftex]{hyperref}
\DeclareGraphicsExtensions{.eps,.bmp,.wmf,.jpg,.pdf}
\numberwithin{equation}{section}
\def\be{\begin{equation}}
\def\ee{\end{equation}}

\def\bea{\begin{eqnarray}}
\def\eea{\end{eqnarray}}

\title{Holographic unification of dark matter and dark energy}
\author{L.N. Granda\thanks{ngranda@univalle.edu.co, ngranda@um.es}\\{\small\it Departamento de Fisica, Universidad del Valle, A.A. 25360 Cali, Colombia}\\
{\small\it Departamento de Fisica, Universidad de Murcia, 30100 Murcia, Spain}}

\date{}
\begin{document}
\maketitle

\begin{abstract}
\noindent 
Using a new version of the holographic principle, a constant term was introduced, which conduces to the description of the standard cosmological $\Lambda$CDM model, and unifies under the same concept the dark matter and dark energy phenomena. The obtained model improves the results of previously considered holographic models based on local quantities. 
The inclusion of constant term is interpreted as a natural first approximation for the infrared cutoff which is associated with the vacuum energy, and the additional terms guarantee an appropriate evolutionary scenario that fits the astrophysical observations. The model allows to reproduce the standard $\Lambda$CDM model without explicitly introducing matter content, and using only geometrical quantities. It is also obtained that the model may describe the dark energy beyond the standard $\Lambda$CDM.
\end{abstract}

\section{Introduction}
\noindent 
The astrophysical data from distant Ia supernovae observations \cite{riess}, cosmic microwave background anisotropy \cite{spergel}, and large scale galaxy surveys \cite{tegmark}, all indicate that the current Universe is not only expanding, it is accelerating due to some kind of  negative-pressure form of matter known as dark energy (\cite{peebles}). The combined analysis of cosmological observations also suggests that the universe is spatially flat, and  consists of about $\sim 1/3$
of dark matter (the known baryonic and nonbaryonic dark matter), distributed in clustered structures (galaxies, clusters of galaxies, etc.) and $\sim 2/3$ of homogeneously distributed (unclustered) dark energy with negative pressure. Despite the high percentage of the dark energy component, its nature as well as its cosmological origin remain unknown at present and a wide variety of models have been proposed to explain the nature of the dark energy and the accelerated expansion (see \cite{peebles} for review).
Among the different models of dark energy, the holographic dark energy approach is quite interesting as it incorporates some concepts of the quantum gravity known as the holographic principle (\cite{bekenstein, thooft, bousso, cohen, susskind}),which first appeared in the context of black holes \cite{thooft} and later extended by Susskind \cite{susskind} to string theory. According to the holographic principle, the entropy of a system scales not with its volume, but with its surface area. In the cosmological context, the holographic principle will set an upper bound on the entropy of the universe \cite{fischler}. In the work \cite{cohen}, it was suggested that in quantum field theory a short distance cut-off $\Lambda$ is related to a long distance cut-off (infra-red cut-off $L$) due to the limit set by black hole formation, namely, if is the quantum zero-point energy density caused by a short distance (UV) cut-off, the total energy in a region of size $L$ should not exceed the mass of a black hole of the same size, thus $L^3\Lambda^4\leq LM_p^2$. Applied to the dark energy issue, if we take the whole universe into account, then the vacuum energy related to this holographic principle is viewed as dark energy, usually called holographic dark energy \cite{cohen} \cite{hsu}, \cite{li}. The largest $L$ allowed is the one saturating this inequality so that the holographic dark energy density is defined by the equality $\rho_\Lambda=3c^2M_p^2L^{-2}$, where $c^2$ is a numerical constant and $M_p^{-2}=8\pi G$.\\
As is well known, the Hubble horizon $H^{-1}$ as the infrared cut-off, gives an equation of state parameter (EoS) equal to zero, behaving as pressureless matter which cannot give accelerated expansion, and the particle horizon gives an EoS parameter larger than $-1/3$, which is not enough to satisfy the current observational data. The infrared cut-off given by the future event horizon yields the desired result of accelerated expansion with an EoS parameter less than $-1/3$, despite the fact that it is based on non local quantities and has problems with the causality \cite{li}. An holographic dark energy model which is based on local and non local quantities have been considered in \cite{sergei}, \cite{sergei1}. Based mainly on dimensional arguments, in \cite{granda}, \cite{granda2} an infrared cut-off for the holographic density was proposed, which describes the dark energy in good agreement with the astrophysical data, and may explain the cosmic coincidence.\\
Despite the available proposals, the theoretical root of the holographic dark energy is still unknown mostly because our ignorance about the microscopic nature of quantum gravity and the origin of dark energy. This translates into a lack of details about the holographic cut-off, and therefore in the present paper we assume that this cut-off should be encoded in a function that depends on natural cosmological quantities like the Hubble parameter, its time derivative and also should contain a constant term, which represents the first approximation to vacuum energy density.\\
In this holographic principle we introduced a constant term that would act as the first natural approximation to the infrared cut-off for the holographic density, since this term may be identified with a constant vacuum energy density, that adjusts very well to the current observational data.
 (see \cite{horava} for a discussion about the cosmological constant in the context of the holographic principle)\\
We propose an holographic density given by a general function of the form $f(H,\dot{H})$, where $f$ should be expanded in powers of his arguments, based on their smallness (compared to the squared Planck mass $M_p^2$). The coefficients of the different terms of this expansion must be found by fitting the model with current observations. At the end we found that this density acquires an interesting geometrical interpretation.\\
As will be shown this approach is able to reproduce the standard $\Lambda$CDM without the necessity of previously introducing any matter term, giving in this way a geometrical interpretation to the most simple and studied cosmological model. Additionally, the results go beyond the standard $\Lambda$CDM model.

\section{The energy density and holographic principle}
The main inequality discussed above $L^3\Lambda^4 \leq LM_p^2$, leads to the restriction involving the UV and IR cut-offs $\Lambda^4\leq M_p^2 L^{-2}$.  Saturating this inequality we come to the concept of holographic dark energy $\rho_{\Lambda}$. To define the infrared cut-off and hence the holographic dark energy density, we propose for $L^{-2}$ a general function which should depend on the main cosmological parameters $H$ and $\dot{H}$, and is measured in units of $(length)^{-2}$, defining the holographic density as follows (from now on we will set $8\pi G=M_p^{-2}=1$)
\begin{equation}\label{eq1}
\rho_\Lambda=3f(H^2,\dot{H})
\ee
where $H=\dot{a}/a$ is the Hubble parameter, and the function $f$ can be Taylor expanded in powers of his arguments provided $H^2$ and $\dot{H}$ are at the present very small quantities measured in Planck mass units, giving
\be\label{eq3}
\rho_\Lambda=3(\lambda+\alpha H^2+\beta \dot{H})
\ee
where we neglected higher order powers of $H^2$ and $\dot{H}$, and $\lambda$, $\alpha$ and $\beta$ are constants to be determined. This model generalizes the one presented in (\cite{granda},\cite{granda2}). The novelty of the present proposal takes root in the presence of the constant term $\lambda$ which we consider as the first natural approximation to the infrared cut-off ($L\sim \lambda^{-1/2}$), and should be associated with the contribution of the constant energy density of the vacuum. As we shall see bellow, the cosmological dynamics depends on $\alpha$ and $\beta$ only through the combination $(\alpha-1)/\beta$ (i.e. cosmological observations fix only this combination) leaving one free parameter. To fix the free parameter we can use geometrical criteria in order to reduce the last two terms in (\ref{eq3}) to a term proportional to Ricci scalar, but the dynamics is independent of the election of this free parameter. \\
\noindent The Friedmann equation in absence of matter is
\begin{equation}\label{eq5}
H^2=\frac{1}{3}\rho_\Lambda=\lambda+\alpha H^2+\beta \dot{H}
\end{equation}
\noindent Setting $x=\ln{a}$, we can rewrite (\ref{eq5}) as follows
\begin{equation}\label{eq6}
H^2=\lambda+\alpha H^2+\frac{\beta}{2}\frac{dH^2}{dx}
\end{equation}
\noindent Introducing the scaled Hubble expansion rate $\tilde{H}=H/H_0$, where $H_0$ is the present value of the Hubble parameter (for $x=0$), the above Friedman equation becomes
\begin{equation}\label{eq7}
\tilde{H}^2=\Omega_{\Lambda}+\alpha\tilde{H}^2
+\frac{\beta}{2}\frac{d\tilde{H}^2}{dx}
\end{equation}

where $\Omega_{\Lambda}=\lambda/H_0^2$. Solving the equation (\ref{eq7}), we obtain
\begin{equation}\label{eq8}
\tilde{H}^2=\Omega_{\Lambda}+\Omega e^{-2x(\alpha-1)/\beta}=\Omega_{\Lambda}+\Omega(1+z)^{2(\alpha-1)/\beta}
\end{equation}
where $\Omega_{\Lambda}=\Omega_{\lambda}/(1-\alpha)$, $\Omega$ is an integration constant and the last equality is written in the redshift variable (by using $e^{-x}=1+z$). 
Note that even without the matter component, this solution is interesting enough as it contains the cosmological constant, and the second term may give matter-like behavior if the constants $\alpha$ and $\beta$ satisfy the restriction $(\alpha-1)/\beta=3/2$, giving rise to the $\Lambda$CDM model. Note also that as the Hubble parameter depends on the combination $(\alpha-1)/\beta$, then cosmological observations can only constraint this combination. Introducing the parameter $\gamma=(\alpha-1)/\beta$ and using the flatness condition $\Omega_{\Lambda}+\Omega=1$, we write the solution (\ref{eq8}) as
\begin{equation}\label{eq8a}
\tilde{H}^2=1-\Omega+\Omega (1+z)^{2\gamma}
\end{equation}
Replacing $\alpha=\gamma\beta+1$ in (\ref{eq3}), we obtain
\be\label{eq8b}
\rho_{\Lambda}=3\left[\lambda+\beta\left(\frac{\beta\gamma+1}{\beta}H^2+\dot{H}\right)\right]
\ee
Once $\gamma$ is fixed by observations, there is still a free parameter $\beta$. To fix  $\beta$ we use geometrical criteria, assuming that the last two terms in (\ref{eq8b}) are proportional to the Ricci scalar. This automatically defines $\beta$ if $(\beta\gamma+1)/\beta=2$. In this case $\beta=1/(2-\gamma)$ and the final expression for the holographic density becomes (the current cosmological observations, and the accelerated expansion of the universe guarantee that $\gamma\neq 2$)
\be\label{eq9}
\rho_{\Lambda}=3\left(\lambda+\frac{1}{2-\gamma}R\right)
\ee
where $R$ is the Ricci curvature. The $\Lambda$CDM model is obtained for $\gamma=3/2$, and the exact holographic density giving rise to the The $\Lambda$CDM is given by
\be\label{eq10}
\rho_{\Lambda CDM}=3\left(\lambda+\frac{1}{3}R\right)
\ee
So it is interesting that the coefficient $1/3$ in (\ref{eq10}) gives rise to $\Lambda$CDM.
Note that regardless of the value of $\gamma$ there is an interesting limit: at far future in the limit $z\rightarrow -1$ the EoS parameter reaches the value $\text{w}=-1$, allowing the possibility that the current universe at the far future will enter in a de Sitter phase.
The present model could be considered as only dark energy model, in which case the free parameter $\gamma$ (to be constrained with observations) allows the possibility of crossing the phantom barrier, describing quintom behavior (see \cite{granda5}).
\section{The phantom barrier}
According to solution (\ref{eq8}), the dark energy component of the holographic density is described by the constant term in (\ref{eq8a}), then it gives automatically the constant DE EoS parameter $\text{w}_0=-1$. From \ref{eq8} it follows the expression for the EoS parameter
\be\label{eq11}
\text{w}(z)=-1+\frac{1}{3}\frac{2\gamma\Omega(1+z)^{2\gamma}}{\Omega_{\Lambda}+\Omega(1+z)^{2\gamma}}
\ee
Note that regardless of the value of $\gamma$ there is an interesting limit: at far future in the limit $z\rightarrow -1$ the EoS parameter reaches the value $\text{w}=-1$, allowing the possibility that the current universe at the far future will enter in a de Sitter phase.
At low redshift we can discuss an interesting approximation. If we argue that in the current universe the power $2\gamma$ in (\ref{eq8a}) differs by a little quantity from $2\gamma=3$, then we can set $2\gamma=3+\delta$ and the Eq. (\ref{eq8a}) becomes
\begin{equation}\label{eq12}
\tilde{H}^2=1-\Omega+\Omega (1+z)^{3+\delta}
\end{equation}
where $\delta$ should be constrained by observations. Expanding the expression (\ref{eq12}) in powers of $\delta$, up to linear order we obtain
\be\label{eq13}
\tilde{H}^2=1-\Omega+\Omega (1+z)^{3}+\Omega\delta (1+z)^3\log(1+z)
\end{equation}
this expansion is valid for $z$ up to $z=e-1$, in order to guarantee the convergence of the expansion.
Separating the matter term in (\ref{eq13}), one defines the dark energy density component as
\be\label{eq14}
\rho_{DE}=3\left(1-\Omega+\Omega\delta (1+z)^3\log(1+z)\right)
\ee
Assuming that the dark matter content obeys the continuity equation, then from the Einstein's equations follows that the DE component automatically satisfies the continuity equation, and we can define the following DE EoS parameter
\be\label{eq15}
\text{w}_{DE}=-1+\frac{\Omega\delta\left[3(1+z)^3\log(1+z)+(1+z)^3\right]}{3\left[1-\Omega+\Omega\delta(1+z)^3\log(1+z)\right]}
\ee
which at $z=0$ gives
\be\label{eq16}
\text{w}_{DE}=-1+\frac{\Omega\delta}{3\left(1-\Omega\right)}
\ee
Note that if $\delta<0$, then the DE would cross the phantom barrier which is also favored by the current astrophysical observations at low redshift. Positive values of $\delta$ give a current EoS parameter $\text{w}_{DE_0}>-1$, that is still within the experimental limits. 
On the other hand, the present model (\ref{eq8a}) could be considered as only dark energy model, in which case the free parameter $\gamma$ (to be constrained with observations) allows the possibility of crossing the phantom barrier, describing quintom behavior (see \cite{granda5}).\\
Intregrating the solution to the Hubble parameter given by Eq. (\ref{eq8a}) we obtain the known time dependence of the scale parameter $a(t)$
\be\label{eq17}
a(t)=\left(\frac{\Omega}{1-\Omega}\right)^{1/3}\left[\sinh\left(\frac{3}{2}\sqrt{1-\Omega}H_0t\right)\right]^{2/3}.
\ee
This solution describes a matter dominated universe at early times ($a(t)\propto t^{2/3}$), and a cosmological constant dominated phase at late times ($a(t)\propto \exp\left(3/2H_0t\right)$).


\section{discussion} We have shown that the holographic principle, may conduce to the description of the fundamental standard model of the current cosmology, and unify under the same concept the dark matter and dark energy phenomena without previously introducing any matter content. The proposed  holographic model, which incorporates a constant term has notorious advantages with respect to the previous models based on local quantities. This holographic principle may also describe a deviation from the $\Lambda$CDM model, allowing the possibility of describing an universe filled with dark energy with varying EoS. 
Adding this constant term to the IR cut-off may improve even models like the Hubble scale which was unable to produce accelerated expansion \cite{hsu}, \cite{li}. In this case if we add a constant term to the squared Hubble scale, we can describe the $\Lambda$CDM model (in this case we should add matter content). 
It was shown that the astrophysical data fixes the combination $\gamma=(\alpha-1)/\beta$. This means that one of the parameters $\alpha$ or $\beta$ still free. To fix the remaining parameter we use geometrical criteria in order to convert the last two terms in (\ref{eq5}) in a term proportional to the Ricci scalar.\\
In short, we can highlight the following important results: The matter and dark energy concepts are unified in a simple geometrical way. The introduction of a cosmological constant term in the frame of the holographic principle attributes quantum gravity nature to the cosmological constant. The holographic principle in the form presented here is able to explain using only geometrical sources, the total matter content of the universe.\\
The holographic principle as presented here induces the following argumentation about the cosmological constant. If we assume that the Einstein-Hilbert action is sufficient to describe the late time cosmological dynamics, then we can put the cosmological constant as a source term in the Einstein equations and conjecture that the most natural choice for the cosmological constant is the geometrical construction $\Lambda(R)=\lambda+\eta R$. For $\eta=1/3$ this conjecture describes in an exact way the $\Lambda$CDM model. And indeed, the $\Lambda$CDM gives a great support for this behavior of the cosmological constant. May be that the right place for the cosmological constant is not in the Einstein-Hilbert action, but in the Einstein equations?. 
\section*{Acknowledgments}
This work was partially supported by Fundacion SENECA (Spain) in the frame of PCTRM 2007-2010.

\end{document}